\documentclass[floatfix,epsf,twocolumn,prb,amsmath,amssymb,showpacs]{revtex4}
\usepackage[T1]{fontenc}
\usepackage{epsfig}
\usepackage{amsmath}
\usepackage{amssymb}
\usepackage{color}
\usepackage{epstopdf}

\newcommand{\p}{\partial}

\begin{document}
\title{Resonant valley filtering of massive Dirac electrons}
\author{D.~Moldovan}\email{dean.moldovan@ua.ac.be}
\author{M.~Ramezani Masir}\email{mrmphys@gmail.com}
\author{L.~Covaci}\email{lucian@covaci.org}
\author{F.~M.~Peeters}\email{francois.peeters@ua.ac.be}
\affiliation{Departement Fysica, Universiteit Antwerpen \\
Groenenborgerlaan 171, B-2020 Antwerpen, Belgium}


\begin{abstract}
Electrons in graphene, in addition to their spin, have two pseudospin degrees of freedom: sublattice and valley pseudospin. Valleytronics uses the valley degree of freedom as a carrier of information similar to the way spintronics uses electron spin. We show how a double barrier structure consisting of electric and vector potentials can be used to filter massive Dirac electrons based on their valley index. We study the resonant transmission through a finite number of barriers and we obtain the energy spectrum of a superlattice consisting of electric and vector potentials. When a mass term is included the energy bands and energy gaps at the K and K' points are different and they can be tuned by changing the potential.
\end{abstract}
\maketitle
\section{Introduction}
The realization of stable single-layer carbon crystals, called graphene, has led to an intensive investigation of graphene's electronic properties \cite{novo1,zhang}. Charge carriers in a sheet of single-layer graphene behave like "relativistic", chiral massless particles with a "light speed" equal to the Fermi velocity and possess a gapless linear spectrum at the $K$ and $K'$ points \cite{zheng,novo1}. During the last decades there have been a lot of theoretical and experimental attempts to use the spin of the electron as a carrier of information\cite{Wolf}. The large research interest in spintronics originates from the fact that it promises to be smaller, more adaptable and faster than today's electronic devices \cite{f1}. Graphene in addition to the spin of the electron has two more degrees of freedom: sublattice and valley pseudospin. Valley-based electronics, also known as valleytronics, uses the valley degree of freedom as a carrier of information similar to the way spintronics uses electron spin.

Valleytronics is useful when the valley encoded information is preserved over long distances. Defects in the hexagonal lattice can cause intervalley scattering and flip the valley state\cite{V1,V2}. This can occur around graphene edges or adatoms that stick to the surface of graphene. However, in order to scatter an electron from the $K$ valley to the $K'$ valley a large transfer of momentum is needed. Typical disorder and Coulomb-type of scattering is unable to provide this momentum and in such a case the valley pseudospin is a conserved quantum number in electronic transport. This allows one to use the valley pseudospin as a carrier of information.

It was shown recently that graphene nanoribbons with zigzag edges\cite{been1,Gar1} can be used as a valley filter. Another promising possibility to control the valley index of electrons is by using a line of defects\cite{mass3}. These can be formed in graphene when grown on a nickel substrate or by using so called mass barriers that can be created by e.g. a proper arrangement of dopants in the graphene sheet\cite{masMass,mass1,mass2}. Another way to control the valley polarization is by using local strain in graphene which induces an effective inhomogeneous magnetic field with opposite sign in the $K$ and $K'$ valleys \cite{A1,A2,A3,A4,A5}. Recently, it was shown that a mass term can be induced by certain substrates such as hexagonal boron nitride (hBN) or by electron-electron interactions which are also able to control the valley pseudospin\cite{Max1,Max2}.

\begin{figure}[ht]
  \centering
  \includegraphics[width=6cm]{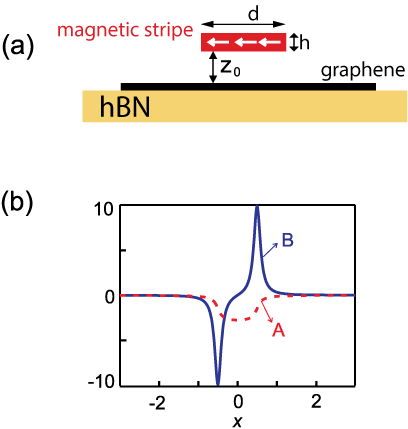}
  \caption{ (Color online) (a) Layout of the system: a ferromagnetic stripe on top of a graphene sheet and a hexagonal boron nitride (hBN) layer as a substrate.
  (b) Perpendicular magnetic field component and corresponding vector potential at a distance $z_{0}$ from the stripe in case of parallel magnetization. }
  \label{First}
\end{figure}
Here we study a double Kronig-Penney (KP) model in graphene, consisting of a  series of electric and vector potential barriers where we include a mass term as induced by e.g. a substrate. The vector potential can be induced using ferromagnetic stripes on top of a graphene layer but such that there is no electrical contact between graphene and these stripes. When one magnetizes the stripes along the $x$ direction, cf. Fig. \ref{First}, by, e.g., applying an in-plane magnetic field, the charge carriers in the graphene layer feel an inhomogeneous magnetic field profile. This profile can be well approximated \cite{mat} by $2B_0z_0h/d(x^2+z_0^2)$  on one edge of the stripe and by $-2B_0z_0h/(x^2+z_0^2)$ on the other, where $z_0$ is the distance between the two-dimensional electron gas (2DEG)  and the stripe, and $d$ and $ h$ the width and height of the stripe (see Fig. \ref{First}). The resulting magnetic field profile will be modeled by two magnetic $\delta$ functions of height $2\pi B_0 h$. Such ferromagnetic stripes have been deposited on top of a 2DEG in a semiconductor heterostructure in Ref. \onlinecite{nog}. This magnetic stripe can also be used as a top gate in order to create an electric potential barrier which will be modeled by a square barrier. These simplified shapes for the electric and magnetic field profiles will allow us to present analytical results. More realistic shapes will not influence the qualitative results of our work.

The model presented in this paper is similar to a recent proposal \cite{Z1} where a single ferromagnetic stripe on top of a graphene layer was shown to act as a valley filter for massive Dirac electrons. In our case the barrier structure consists of two ferromagnetic stripes acting as a resonant tunneling structure. We investigate both symmetric and antisymmetric functions of the electric and vector potentials and we show how different potential symmetries affect the transmission of the electrons in the $K$ and $K'$ valleys. Furthermore, a superlattice consisting of such a barrier structure is considered. We show that the band structure of the superlattice is different for $K$ and $K'$ valleys and most importantly that the height of the band gap is valley dependent and can be tuned by adjusting the potential.

The manuscript is organized as follows. In Sec. II we present the model and analytical tools that were used. In Sec. III we evaluate the transmission for the combined electric and vector potential barrier. In Sec. IV we expand the analysis to include multiple unit cells. In Sec. V we consider a superlattice of such barriers and we evaluate the valley dependent band structure and band gap. Our concluding remarks are given in Sec. VI.
\section{Model}
The Dirac Hamiltonian for a massive electron in the presence of an external electric and vector potential is given by \cite{Beenrev},
\begin{equation} \label{dirac_hamiltonian_base}
   \mathcal{H} = V + \tau m v_F^2 \sigma_z + v_F({\bf{p}} + e \textbf{A})\cdot\boldsymbol{\sigma},
\end{equation}
where ${\bf{p}}  = -i\hbar(\p_x, \p_y)$ is the momentum operator, $\sigma_z$ and $\boldsymbol{\sigma} = (\sigma_x, \sigma_y)$ are the Pauli matrices and  $\tau = \pm 1$ is the valley index. The behavior of the electron is described by the equation,
\begin{equation} \label{dirac_equation_base1}
  \mathcal{H}\Psi_\tau = E\Psi_\tau.
\end{equation}
The equation can be simplified by using dimensionless units via the following substitutions: $W \rightarrow lW$, $l = 1$ nm, $E \rightarrow E_0E$, $V \rightarrow E_0 V$, $E_0 = \hbar v_F / l \approx 658$ meV, $A(x) \rightarrow (B_0l/\beta) A(x)$,  $\beta = eB_0 l^2 / \hbar$, $\mu = mv_F^2 / E_0$. We choose the Landau gauge in which ${\bf{A}} = (0, A(x))$. Since $p_{y}$ commutes with the Hamiltonian of Eq. (\ref{dirac_hamiltonian_base}) then $p_y$ is a good quantum number, and due to the translational invariance along the $y$ direction the solutions have the form,
\begin{equation}\label{Wavexy}
  \Psi_\tau(x,y) = e^{ik_y y} \psi_\tau(x).
\end{equation}
After substituting Eq. (\ref{Wavexy}) into Eq. (\ref{dirac_equation_base1}), we find,
\begin{equation}
  \label{dirac_equation_dimensionless_y}
  \begin{pmatrix}
    V + \tau\mu & -i\p_x - i(k_y + A) \\
    -i\p_x + i(k_y + A) & V - \tau\mu
  \end{pmatrix}
  \psi_\tau = E\psi_\tau.
\end{equation}
In order to find the energy spectrum for the $K$ valley we use $\tau \rightarrow 1$ and $k_y \rightarrow +k_y$, while for the $K'$ valley we replace $\tau \rightarrow -1$ and $k_y \rightarrow -k_y$.

\section{Barrier structure}
Let us first consider a single barrier consisting of both electric and vector potentials given by,
\begin{eqnarray} \label{potential_triple}
  V(x)= V_b \left[ \Theta(x) - \Theta(x-W_b) \right], \\
  A(x)= A_b \left[ \Theta(x) - \Theta(x-W_b) \right],
\end{eqnarray}
where $\Theta$ is the step function, $V_b$ and $A_b$ are constant values of the electric and vector potentials and $W_b$ is the width of the barrier.

The incident wave function before the barrier is given by,
\begin{equation}
  \psi_{in}(x) =
  \begin{pmatrix}
    e^{ik_0 x} + r e^{-ik_0 x} \\
    f_{0} e^{i(k_0 x + \theta_0)} - r f_{0} e^{-i(k_0 x + \theta_0)}
  \end{pmatrix},
\end{equation}
and the transmitted wave function after the barrier is,
\begin{equation}
  \psi_{out}(x) =
  \begin{pmatrix}
    t e^{ik_0 x} \\
    t f_{0} e^{i(k_0 x + \theta_0)}
  \end{pmatrix}.
\end{equation}
where $k_0 = \sqrt{E^2 - \mu^2 - k_y^2}$ is the wave vector, with $\theta_0 = \arctan(k_y/k_0)$ and $f_0 = {\sqrt{E^2 - \mu^2}}/{(E + \tau \mu)}$. For a zero mass term $\mu = 0$ the $f_0$ component reduces to $sgn(E)$.

Inside the barrier we have the wave function,
\begin{equation} \label{wavefunction}
  \psi_b(x) =
  \begin{pmatrix}
    A e^{ik_{b} x} + B e^{-ik_{b} x} \\
    A f_b e^{i(k_{b} x + \theta_b)} - B f_b e^{-i(k_{b} x + \theta_b)}
  \end{pmatrix},
\end{equation}
where $k_b = \sqrt{(E-V_b)^2 - \mu^2 - (k_y+A_b)^2}$ is the wave vector, with the angle $\theta_b = \arctan[(k_y+A_b)/k_b]$ and the component $f_b = {\sqrt{(E-V_b)^2 - \mu^2}}/{(E-V_b + \tau \mu)}$.

By requiring the continuity of the wave function at the barrier boundaries, we find the relation between the wave function coefficients before and after the barrier. This relation can be expressed as a $2\times2$ transfer matrix,
\begin{equation}
    \begin{pmatrix}
        1 \\
        r
    \end{pmatrix}
     = M
    \begin{pmatrix}
        t \\
        0
    \end{pmatrix},
\end{equation}
where the matrix elements are given by,
\begin{widetext}
    \begin{equation} \label{M11}
        M_{11} = M_{22}^* = e^{ik_0W_b} \left[ \cos(k_bW_b) + i\sin(k_bW_b) \left( \frac{\sin\theta_0\sin\theta_b - \lambda}{\cos\theta_0\cos\theta_b} \right) \right],
    \end{equation}
    \begin{equation} \label{M12}
        M_{12} = M_{21}^* = \frac{e^{-ik_0W_b}e^{-i\theta_0}\sin(k_bW_b)}{\cos\theta_0\cos\theta_b} \bigg[ \lambda\sin\theta_0 - \sin\theta_b + i\tau\nu \cos\theta_0 \bigg],
    \end{equation}
\end{widetext}
where,
\begin{equation}
    \lambda = \frac{E(E-V_b) - \mu^2}{\sqrt{E^2 - \mu^2}\sqrt{(E-V_b)^2 - \mu^2}},
\end{equation}
\begin{equation} \label{gamma_i}
    \nu = \frac{\mu V_b}{\sqrt{E^2 - \mu^2}\sqrt{(E-V_b)^2 - \mu^2}}.
\end{equation}
The transmission and reflection coefficients can be found from the transfer matrix elements as $t=1/M_{11}$ and $r=M_{12}/M_{11}$.

Notice that, unlike $M_{11}$, element $M_{12}$ is valley dependent. To investigate further, we can rewrite the complex elements of the transfer matrix to separate the amplitude and phase as $M_{11} = m_{11} e^{i\varphi_{11}}$, $M_{12} = m_{12} e^{i\varphi_{12}}$. For $M_{11}$ we find,
\begin{equation}
    m_{11} = \sqrt{ \cos^2(k_b W_b) + \sin^2(k_b W_b) \left( \frac{\sin\theta_0\sin\theta_b - \lambda}{\cos\theta_0\cos\theta_b} \right)},
\end{equation}
\begin{equation} \label{varphi_i11}
    \varphi_{11} = k_0W_b + \arctan\left( \frac{\sin\theta_0\sin\theta_b - \lambda}{\cos\theta_0\cos\theta_b} \tan(k_b W_b) \right).
\end{equation}
In this case the transmission coefficient can be written as $t = |t|e^{i\varphi_t}$. Both the transmission amplitude $|t|=1/m_{11}$ and phase $\varphi_t = -\varphi_{11}$ are independent of $\tau$. Next we look at $M_{12}$,
\begin{equation}
    m_{12} = \frac{\sin(k_bW_b)}{\cos\theta_0\cos\theta_b} \sqrt{ (\lambda\sin\theta_0 - \sin\theta_b)^2 + \nu^2 \cos^2\theta_0 },
\end{equation}
\begin{equation} \label{varphi_i12}
    \varphi_{12} = -k_0W_b - \theta_0 + \arctan\left( \frac{\tau \nu \cos\theta_0}{\lambda\sin\theta_0 - \sin\theta_b} \right).
\end{equation}
The reflection coefficient can be written as $r=|r|e^{i\varphi_r}$, where $|r| = m_{12}/m_{11}$ and $\varphi_r = \varphi_{12} - \varphi_{11}$. The amplitude $m_{12}$ no longer depends on $\tau$ because $\tau^2=1$. This means that the reflection probability $R = |r|^2$ is also independent of the valley index. However, the phase $\varphi_{12}$ is still valley dependent.

Electrons that are reflected at this barrier will have a slight phase difference depending on their valley index. This does not in any way affect the transmission and reflection probabilities for the single barrier, but, as we will see later, it has a significant impact on the resonance peaks of a double barrier structure.

\begin{figure}[h]
  \centering
  \includegraphics[width=5.9cm]{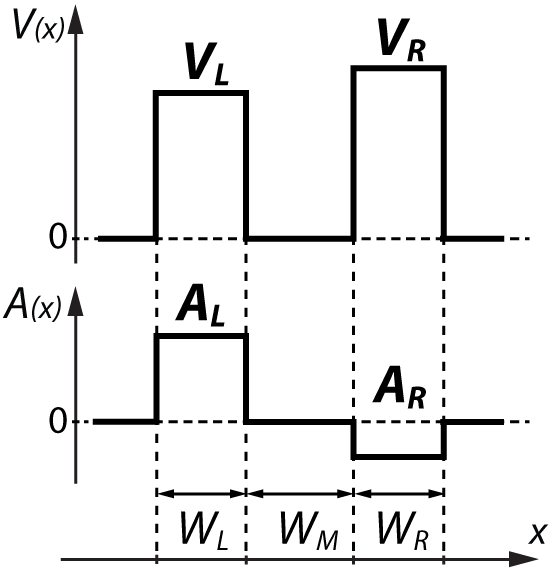}
  \caption{(Color online) Double barrier structure with electric and vector potentials.}
  \label{fig:B3_potential}
\end{figure}

We now consider a structure that consists of two barriers as shown in Fig. \ref{fig:B3_potential}. The left (right) barrier consists of an electric potential $V_L$ ($V_R$) and a vector potential $A_L$ ($A_R$). The two barriers are separated by a distance $W_M$.
\begin{figure}[ht]
  \centering
  \includegraphics[width=8.6cm]{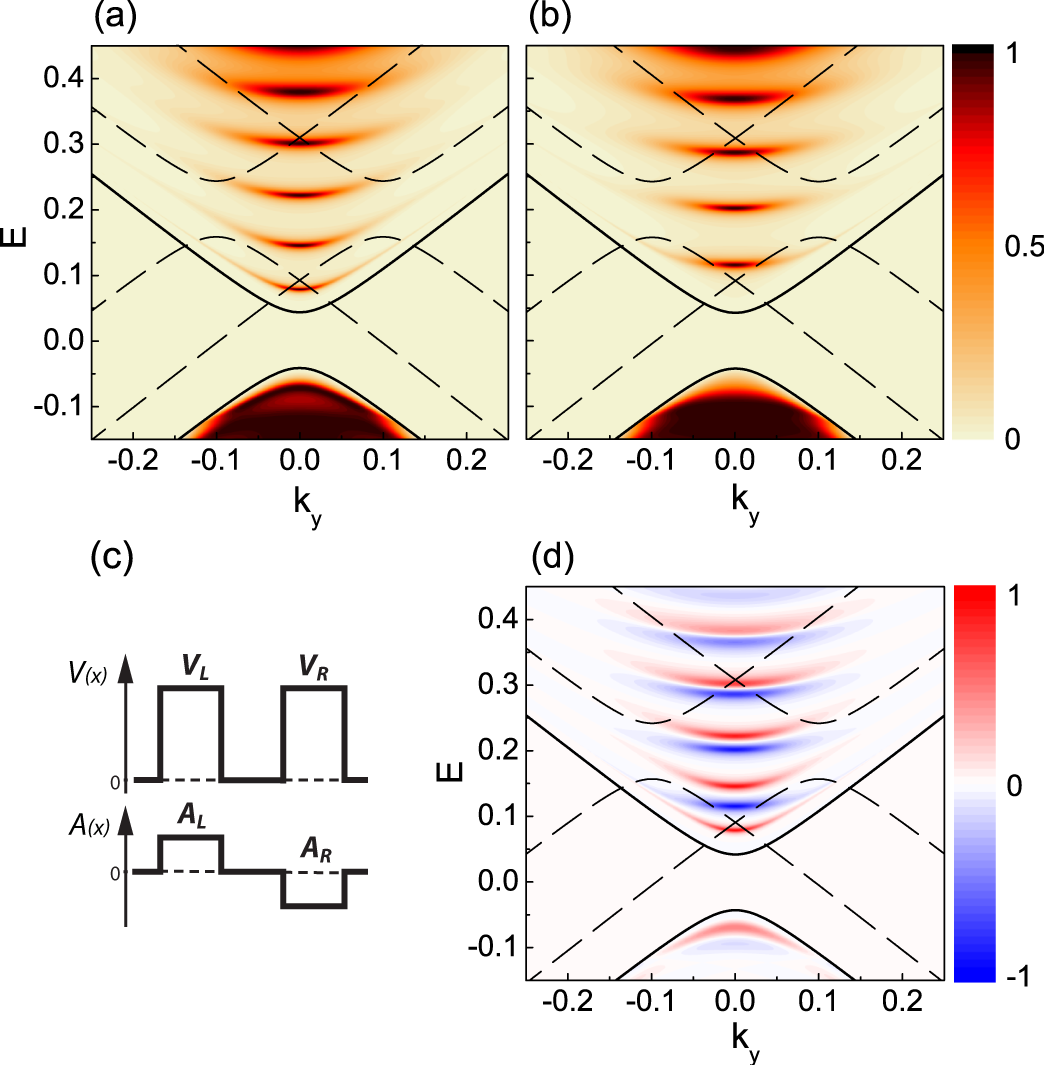}
  \caption{(Color online) Transmission probability for the (a) $K$ and (b) $K'$ valleys. (c) The barrier structure consists of a symmetric electric potential and an antisymmetric vector potential. (d) The difference $T_K - T_{K'}$. The solid black line shows the energy dispersion of a free massive electron in bulk graphene, $E=\pm\sqrt{\mu^2+k_y^2}$, and the dashed lines show the dispersion inside the barriers, $E=V_i \pm \sqrt{\mu^2+(k_y+A_i)^2}$. The parameters are: $V_L=V_R=0.2$, $A_L=-A_R=0.1$, $\mu=0.0425$, $W_L=W_R=10$, $W_M=30$.}
  \label{fig:B4_transmission_const_V}
\end{figure}
\begin{figure}[ht]
  \centering
  \includegraphics[width=8.6cm]{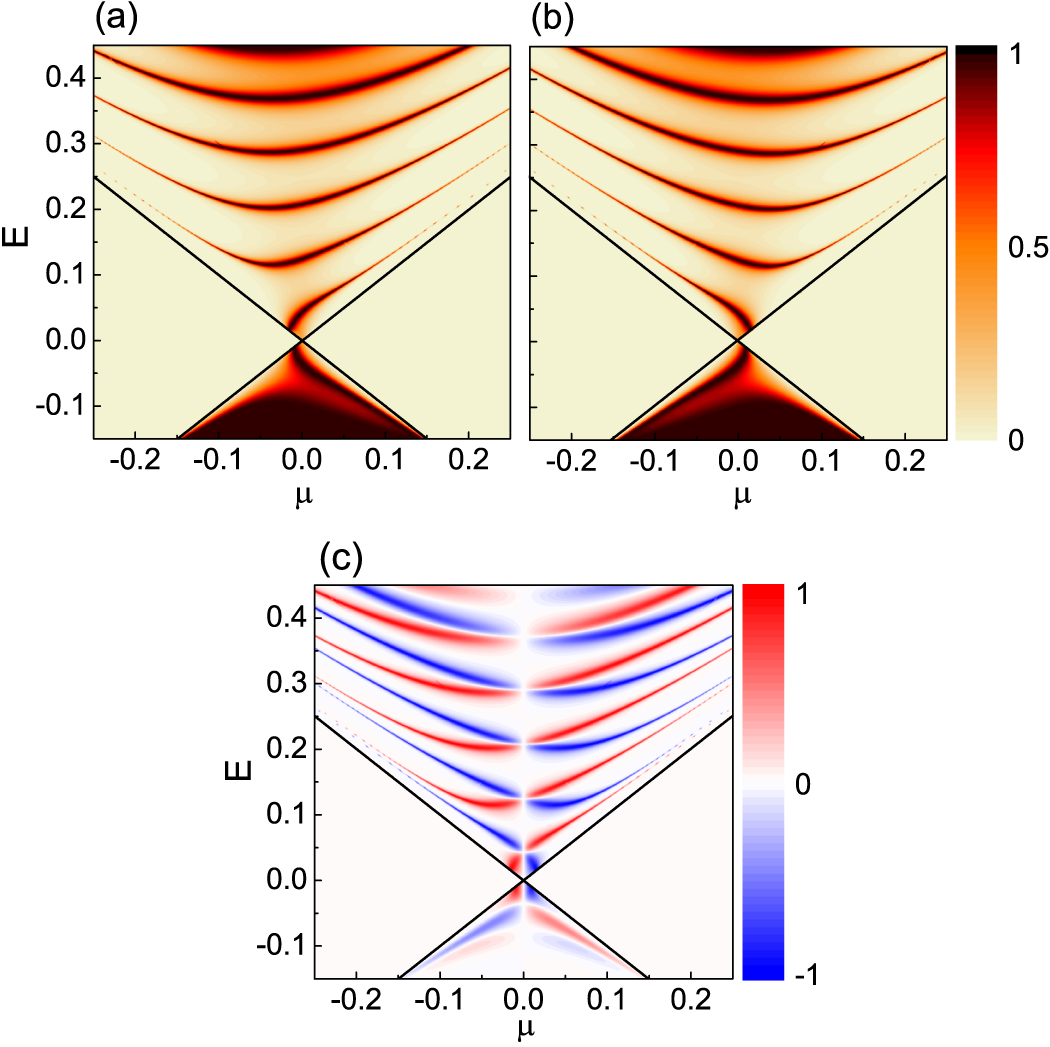}
  \caption{(Color online) Transmission probability at $k_y=0$ as a function of the mass term $\mu$ and the energy for the (a) $K$ and (b) $K'$ valleys. (c) The difference $T_K - T_{K'}$. The solid black line shows $E=\mu$. The barrier structure is the same as in Fig. \ref{fig:B4_transmission_const_V}.}
  \label{fig:B4_varM_const_V}
\end{figure}
\begin{figure}[h]
  \centering
  \includegraphics[width=8.0cm]{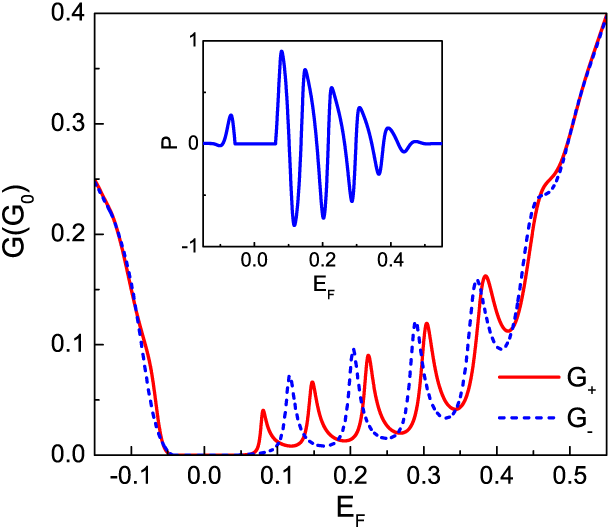}
  \caption{(Color online) Total zero-temperature conductance as a function of Fermi energy for the same barrier structure as in Fig. \ref{fig:B4_transmission_const_V}. The inset shows the valley polarization. }
  \label{fig:B4_conductance_const_V}
\end{figure}
\begin{figure}[h]
  \centering
  \includegraphics[width=8.0cm]{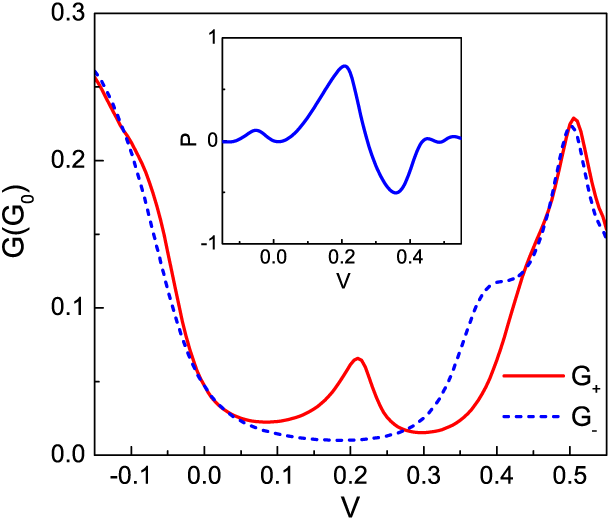}
  \caption{(Color online) Total zero-temperature conductance as a function of the electric potential at Fermi energy $E_F=0.15$. The barrier structure is the same as in Fig. \ref{fig:B4_transmission_const_V}, but with a variable electric potential $V_L=V_R=V$. The inset shows the valley polarization. }
  \label{fig:B4_conductance_varV_const_V}
\end{figure}

The total transfer matrix $M_T$ for the double barrier structure is defined as,
\begin{equation} \label{compMat}
   \begin{pmatrix}
     1 \\
     r
  \end{pmatrix}
  = M_T
  \begin{pmatrix}
     t \\
     0
  \end{pmatrix}
  = M_L M_M M_R
  \begin{pmatrix}
     t \\
     0
  \end{pmatrix},
\end{equation}
where $M_L$ and $M_R$ are the transfer matrices of the left and right barrier, respectively, and $M_M$ models the barrier spacing. The transfer matrix elements for the individual barriers are the same as the previously derived elements in Eqs. (\ref{M11}) and (\ref{M12}) with the barrier index replaced as $b\rightarrow L$ and $b\rightarrow R$ for the left and right barrier, respectively. For the middle region we define the propagation matrix $M_{M}$ to connect the coefficients of the two barriers,
\begin{equation}
   M_M =
   \begin{pmatrix}
     e^{-ik_0W_M} & 0 \\
     0 & e^{ik_0W_M}
  \end{pmatrix}.
\end{equation}
Transmission coefficient of the double barrier structure is given by $t=1/M_{T11}$, so we only need to find the $M_{T11}$ element of the total transfer matrix from Eq. (\ref{compMat}),
\begin{equation} \label{M_11_total}
   M_{11} = M_{L11} M_{R11}e^{-ikW_M} + M_{L12} M_{R21}e^{ikW_M}.
\end{equation}
We can substitute the complex matrix elements as $M_{ij} = m_{ij} e^{i\phi_{ij}}$ in Eq. (\ref{M_11_total}) and derive the total transmission probability,
\begin{equation} \label{transmission_RTS}
    T = \frac{1}{|M_{11}|^2} = \frac{ T_L T_R }{ 1 + R_L R_R + 2\sqrt{R_L R_R} \cos{\phi_\tau} },
\end{equation}
where $T_L$ ($T_R$) and $R_L$ ($R_R$) are the transmission and reflection probabilities of the left (right) barrier, respectively, and $\phi_\tau$ is the total phase factor,
\begin{equation} \label{transmission_phase}
    \phi_\tau = 2k_{0} W_M - \varphi_{L11} - \varphi_{R11} + \varphi_{L12} - \varphi_{R12},
\end{equation}
where $\varphi_{L11}$, $\varphi_{L12}$ ($\varphi_{R11}$, $\varphi_{R12}$) are the phase factors of the left (right) barrier given by Eqs. (\ref{varphi_i11}) and (\ref{varphi_i12}). Because of the phase factor $\phi_\tau$ resonant transmission peaks will occur when,
\begin{equation} \label{transmission_phase_max}
    \phi_\tau = (2n+1)\pi,\qquad n = 0,1,....
\end{equation}
The transmission maximum is thus,
\begin{equation}
    T_{max} = \frac{ T_L T_R }{ (1 - \sqrt{R_L R_R})^2 }.
\end{equation}
If the transmission probabilities $T_{R}$ and $T_{L}$ are small (as they are here), the reflection probabilities in the denominator may be expanded as,
\begin{equation}
   1 - \sqrt{R_L R_R} \approx 1 - \left(1-\frac{T_{L}}{2} \right) \left( 1-\frac{T_{R}}{2} \right) \approx \frac{T_{L} + T_{R}}{2},
\end{equation}
and the total transmission probability is,
\begin{equation}
    T_{max} \approx \frac{ 4T_L T_R }{ (T_{L} + T_{R})^2 }.
\end{equation}
The minimum transmission of resonance occurs when the cosine function in Eq. (\ref{transmission_RTS}) is unity, i.e. when $\phi_\tau = 2 n \pi$.

The single barrier transmission and reflection probabilities for the left and right barrier are completely independent of the valley index. As shown earlier, the phase $\varphi_{12}$ is valley dependent and as a consequence the resonant peaks of the double barrier structure will occur at different energies for the two valleys.
\begin{figure}[h]
  \centering
  \includegraphics[width=8.6cm]{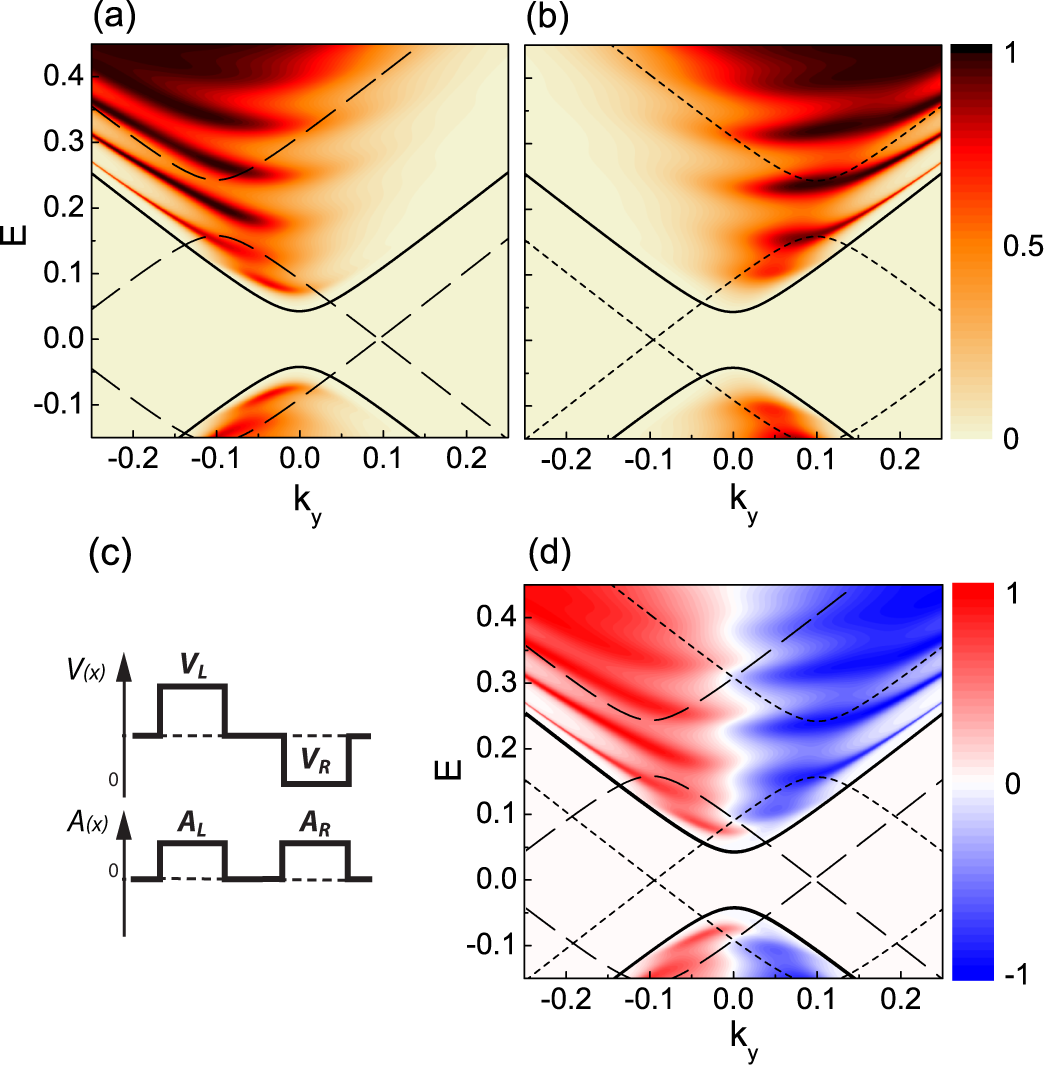}
  \caption{(Color online) Transmission probability for the (a) $K$ and (b) $K'$ valleys. (c) The barrier structure consists of a symmetric vector potential and an antisymmetric electric potential. (d) The difference $T_K - T_{K'}$. The solid black line shows the energy dispersion of a free massive electron in bulk graphene, $E=\pm\sqrt{\mu^2+k_y^2}$, and the dashed lines show the dispersion inside the barriers, $E=V_i \pm \sqrt{\mu^2+(k_y+A_i)^2}$. The parameters are: $V_L=-V_R=0.2$, $A_L=A_R=0.1$, $\mu=0.0425$, $W_L=W_R=10$, $W_M=30$.}
  \label{fig:B4_transmission_const_A}
\end{figure}
\begin{figure}[ht]
  \centering
  \includegraphics[width=8.6cm]{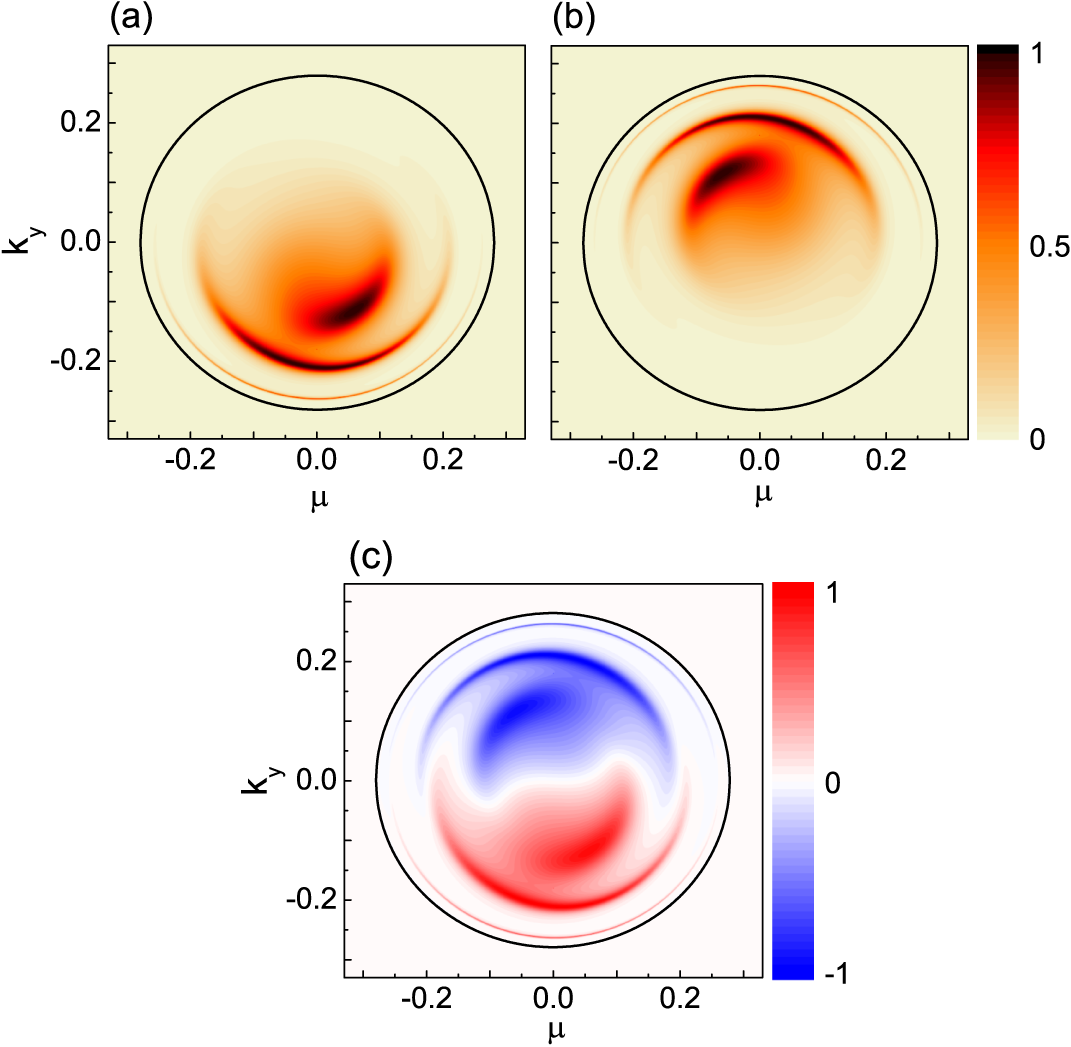}
  \caption{(Color online) Transmission probability at $E=0.28$ as a function of the mass term $\mu$ and $k_y$ for the (a) $K$ and (b) $K'$ valleys. (c) The difference $T_K - T_{K'}$. The solid black line shows $E= \sqrt{\mu^2 + k_y^2}$. The barrier structure is the same as in Fig. \ref{fig:B4_transmission_const_A}.}
  \label{fig:B4_varM_const_A}
\end{figure}
\begin{figure}[ht]
  \centering
  \includegraphics[width=8.0cm]{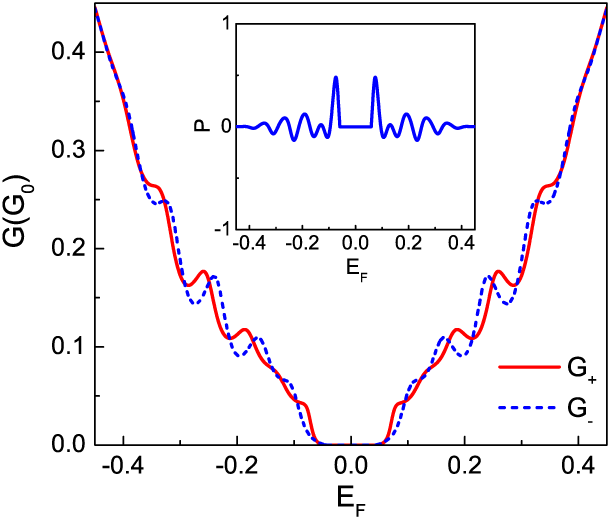}
  \caption{(Color online) Total zero-temperature conductance as a function of Fermi energy for the same barrier structure as in Fig. \ref{fig:B4_transmission_const_A}. The inset shows the valley polarization.}
  \label{fig:B4_conductance_const_A}
\end{figure}
\begin{figure}[h]
  \centering
  \includegraphics[width=8.0cm]{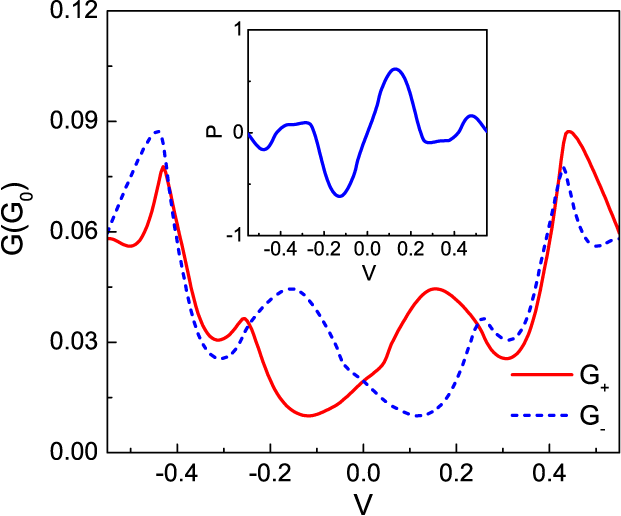}
  \caption{(Color online) Total zero-temperature conductance as a function of the electric potential at Fermi energy $E_F=0.08$. The barrier structure is the same as in Fig. \ref{fig:B4_transmission_const_A}, but with a variable electric potential $V_L=V_R=V$. The inset shows the valley polarization.}
  \label{fig:B4_conductance_varV_const_A}
\end{figure}
\begin{figure}[t]
  \centering
  \includegraphics[width=8.6cm]{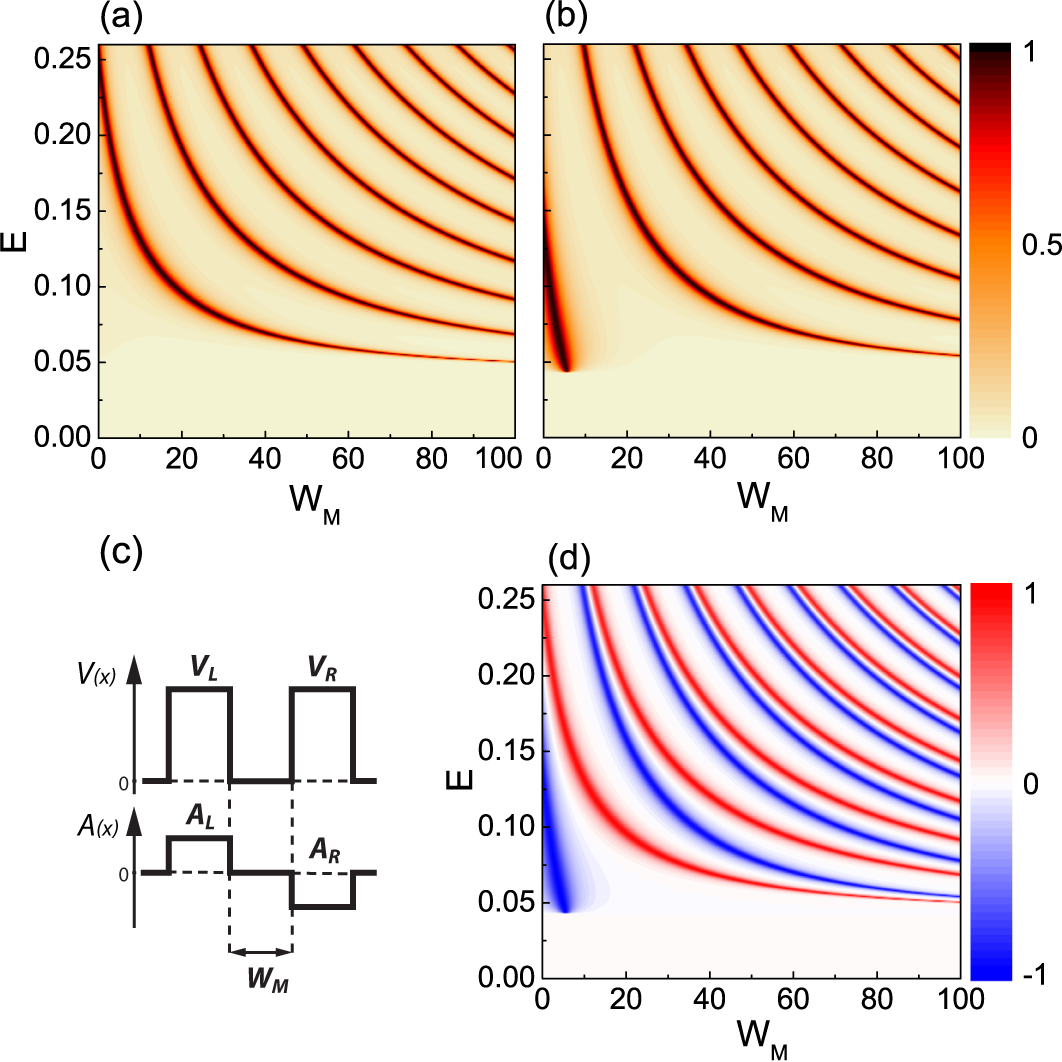}
  \caption{(Color online) Transmission probability at $k_y=0$ as a function of energy and barrier spacing $W_M$ for the (a) $K$ and (b) $K'$ valleys. (c) The barrier structure consists of a symmetric electric potential and an antisymmetric vector potential. (d) The difference $T_K - T_{K'}$. The parameters are: $V_L=V_R=0.2$, $A_L=-A_R=0.1$, $\mu=0.0425$, $W_L=W_R=10$. }
  \label{fig:B4_transmission_Wvar}
\end{figure}

For the valley dependent phase difference $\varphi_{L12} - \varphi_{R12}$ we find,
\begin{equation} \label{phi_tau_difference}
  \varphi_{L12} - \varphi_{R12} = -k_0(W_L-W_R) + \arctan ( \gamma_\tau \omega ),
\end{equation}
where,
\begin{eqnarray} \label{gamma_barrier}
  \gamma_\tau &=& \tau\mu(V_L A_R - V_R A_L), \\
  \omega &=& \frac{ k_0 (\mu^2-E^2) }{ k_0 \mu^2 V_L V_R + Q_L Q_R}, \\
  Q_i &=& EV_i k_y + (E^2-\mu^2)(k_y+A_i).
\end{eqnarray}
The valley index dependent factor is $\gamma_\tau$ and the system must satisfy the condition $\gamma_\tau \neq 0$ in order to show any valley index dependent behavior,
\begin{equation} \label{valley_condition}
  \mu (V_L A_R - V_R A_L) \neq 0.
\end{equation}
This implies that the minimum requirement is a non-zero mass term $\mu$. Furthermore, both an electric and a vector potential are required, but this condition $(V_L A_R - V_R A_L \neq 0)$ is more flexible with regard to the individual potential values. The system must consist of two barriers with different electric and/or vector potentials. In the case of a single barrier, i.e. $V_R = 0$ and $A_R = 0$, condition (\ref{valley_condition}) cannot be fulfilled. If both a non-zero mass term and two barriers are present, the condition reduces to $V_L A_R \neq V_R A_L$. Notice that the condition is still satisfied if one barrier has only an electric potential ($V_L=V$, $A_L=0$), while the other has only a vector potential ($V_R=0$, $A_R=A$).

Considering the result of the phase difference, Eq. (\ref{phi_tau_difference}), we can rewrite Eq. (\ref{transmission_phase}) as,
\begin{equation}
    \phi_\tau = \varphi_0 + \tau\varphi_1,
\end{equation}
where $\varphi_0$ and $\tau\varphi_1$ are the valley index independent and dependent phase factors, respectively,
\begin{eqnarray}
    \varphi_0 &=& 2k_{0} W_M - \varphi_{L11} - \varphi_{R11} - k_0(W_L-W_R), \\
    \varphi_1 &=& sgn(\gamma \omega)|\arctan ( \gamma \omega )|, \\
    \gamma &=& \mu(V_L A_R - V_R A_L).
\end{eqnarray}
This means that the phase factor in the $K$ valley will be $\phi_K = \varphi_0 + \varphi_1$, while the $K'$ valley will have $\phi_{K'} = \varphi_0 - \varphi_1$. Considering that $\cos\phi_\tau$ is a symmetric function, we need to analyze the absolute value $|\phi_\tau|$. The phase difference between $K$ and $K'$ is thus,
\begin{equation}
    \Delta\phi = |\phi_K| - |\phi_{K'}|.
\end{equation}
If $\varphi_0 > \varphi_1$ we find,
\begin{equation}
    \Delta\phi = \varphi_0 + \varphi_1 - (\varphi_0 - \varphi_1) = 2\varphi_1.
\end{equation}
On the other hand if $\varphi_0 < \varphi_1$,
\begin{equation}
    \Delta\phi = \varphi_0 + \varphi_1 - (\varphi_1 - \varphi_0) = 2\varphi_0.
\end{equation}
Because of this, we expect to see two different resonance domains with different peak distributions.

\textbf{\emph{Conductance and polarization:}} For a given Fermi energy $E_F$, the valley-related zero-temperature conductance is calculated from:
\begin{equation}\label{cond}
  G_\tau = G_0 \int_{k_{y-}}^{k_{y+}} T_\tau(E_F, k_y),
\end{equation}
where $G_0 = e^2 L_y/\pi h$ and $k_{y\pm} = \pm \sqrt{E_F^2-\mu^2}$.
The total conductance is $G = G_+ + G_-$, with the valley polarization defined as,
\begin{equation}\label{polar}
  P = \frac{G_+ - G_-}{G}.
\end{equation}

\emph{\textbf{Numerical results:}} The transmission probability is calculated numerically based on Eq. (\ref{transmission_RTS}). We consider two cases. First, as shown in Fig. \ref{fig:B4_transmission_const_V}(c) we take the electric potential of the two barriers as symmetric, while the vector potential is taken antisymmetric. The transmission is shown in Figs. \ref{fig:B4_transmission_const_V}(a) and (b) for $K$ and $K'$, respectively. Transmission resonances are found at different energy values for $K$ and $K'$ as described by Eq. (\ref{transmission_RTS}). In order to emphasize the difference we subtracted the electron transmission in the two valleys and plotted the result in Fig. \ref{fig:B4_transmission_const_V}(d). Notice that the difference is quite large and changes between -1 and 1, which means that for some energy values while the electron in one valley is completely transmitted, for the same energy we have complete reflection in the other valley.

We plotted the transmission as a function of the mass term $\mu$ and energy in Fig. \ref{fig:B4_varM_const_V}. We can see that the two valleys have exactly opposite behavior with regard to the sign of the mass term. As the value of the mass term is increased, the energy values of the resonant peaks are lowered for one valley and increased in the other. A notable exception is the first resonant peak which behaves differently and vanishes into the band gap. This is a special case present in the domain where $\varphi_0 < \varphi_1$ as discussed earlier.

We evaluated the conductance and polarization using Eqs. (\ref{cond}) and (\ref{polar}). The results are shown in Fig. \ref{fig:B4_conductance_const_V} where pronounced valley polarization is found for certain energies. High polarization peaks are present at lower values of Fermi energy, but as the energy is increased, the polarization becomes lower and eventually converges to zero. We also plotted the conductance as a function of the electric potential in Fig. \ref{fig:B4_conductance_varV_const_V}, where both positive and negative polarization peaks are present. This shows that the valley filtering properties can easily be switched by adjusting the electric potential.

Next, we consider the case where the electric potential is antisymmetric, while the vector potential is symmetric as shown in Fig. \ref{fig:B4_transmission_const_A}(c). The transmission in this case is plotted in Figs. \ref{fig:B4_transmission_const_A}(a) and (b). The difference between the two valleys is most apparent when moving from
$k_y < 0$ to $k_y > 0$, but this is expected because of the $k_y \rightarrow -k_y$ substitution when switching the valleys. The more important result are the energy dependent transmission differences. They are a consequence of the valley index dependent resonances that were introduced with the mass term in the double barrier structure. We plotted the transmission as a function of the mass term $\mu$ and $k_y$ at a fixed energy in Fig. \ref{fig:B4_varM_const_A}. As in the previous case, we see that the valleys have opposite behavior with regard to the sign of the mass term.

The conductance and polarization are plotted in Fig. \ref{fig:B4_conductance_const_A}. Compared with the previous case, the conductance is generally larger, but the valley polarization is much less pronounced. Like previously, it converges to zero for high energy values. Conductance is plotted as a function of the electric potential in Fig. \ref{fig:B4_conductance_varV_const_A}. In this case the polarization is antisymmetric with respect to $V=0$, which means the type of valley polarization can be switched by changing the sign of the electric potential.

Finally, it's also possible to realize valley dependent transmission if the left barrier has only an electric potential, while the right barrier has only a vector potential ($V_L = V, V_R = 0, A_L = 0, A_R = A$). However, while interesting, there is no special advantage to this configuration. The first case, with a symmetric electric potential and an antisymmetric vector potential shows the largest energy dependent valley polarization and therefore this is the only configuration that will be discussed from now on.

We also consider the impact of the length of the spacing between barriers $W_M$ on the transmission. Figs. \ref{fig:B4_transmission_Wvar} (a) and (b) show energy values of high transmission for the $K$ and $K'$ valleys and Fig. \ref{fig:B4_transmission_Wvar}(d) shows the difference in transmission between the two valleys.  The first transmission peak of the $K'$ valley in Fig. \ref{fig:B4_transmission_Wvar}(b) vanishes, similar to the behavior shown in Fig. \ref{fig:B4_varM_const_V}(b). In this case the $K'$ valley exhibits resonances from both $\varphi_0 < \varphi_1$ and $\varphi_0 > \varphi_1$ domains, while in the $K$ valley we always have $\varphi_0 > \varphi_1$.

The transmission peaks in Fig. \ref{fig:B4_transmission_Wvar} are described by Eq. (\ref{transmission_phase_max}). For large energy and small $W_M$, their positions are valley dependent and we can approximate them by,
\begin{equation}
    E_n = \sqrt{\mu^2 + \left(\frac{1}{2} E_0 + \frac{1}{2}\sqrt{ E_0^2 + \tau E_\tau} \right)^2},
\end{equation}
where,
\begin{eqnarray}
    E_0 &=& \frac{(2n+1)\pi}{2W_t},\\ 
    E_\tau &=& \frac{2\mu}{W_t}  \left( \frac{V_L}{A_L} - \frac{V_R}{A_R}\right), \\
    W_t &=& W_M + \frac{1}{2} (W_L + W_R).
\end{eqnarray}
On the other hand, if $W_M$ is large, the peak positions are no longer valley dependent and Eq. (\ref{transmission_phase_max}) can be approximated by,
\begin{equation}
    E_n = \sqrt{\mu^2 + \left( \frac{(2n+1)\pi}{2W_M} \right)^2},
\end{equation}
which agrees with the asymptotic behavior shown in Figs. \ref{fig:B4_transmission_Wvar} (a) and (b).

\section{$N$ unit cells}
\begin{figure}[b]
  \centering
  \includegraphics[width=6.5cm]{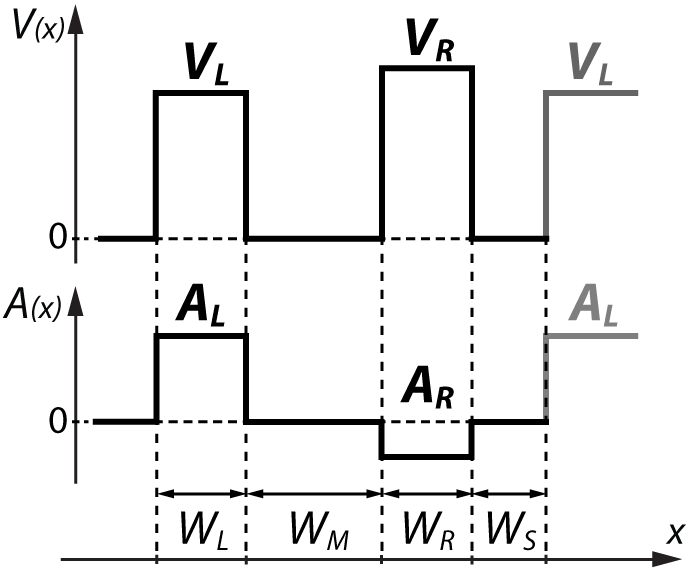}
  \caption{(Color online) Double barrier unit cell with electric and vector potentials.}
  \label{fig:SL4_potential}
\end{figure}
\begin{figure}[t]
  \centering
  \includegraphics[width=8.6cm]{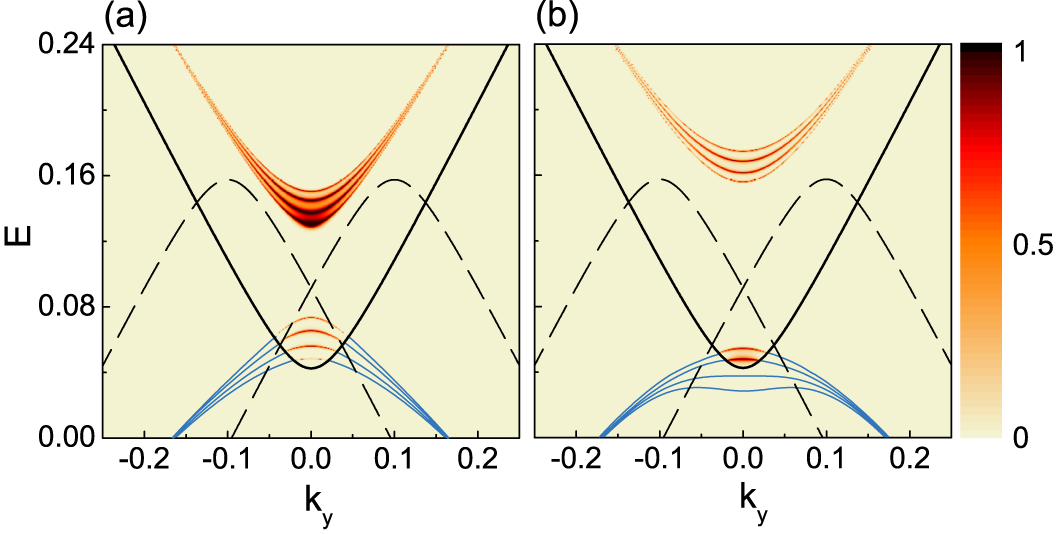}
  \caption{(Color online) Transmission probability and bound states (blue lines) for a system with $5$ unit cells for the (a) $K$ and (b) $K'$ valleys. The solid black line shows the energy dispersion of a free massive electron in bulk graphene, $E=\pm\sqrt{\mu^2+k_y^2}$, and the dashed lines show the dispersion inside the barriers, $E=V_i \pm \sqrt{\mu^2+(k_y+A_i)^2}$. The parameters are: $V_L=V_R=0.2$, $A_L=-A_R=0.1$, $\mu=0.0425$, $W_L=W_R=W_M=10$, $W_S=5$.}
  \label{fig:B4_N5_transmission}
\end{figure}
Next, we repeat the previous barrier structure $N$ times. We add a second spacing region $W_S$ to separate unit cells as shown in Fig. \ref{fig:SL4_potential}. The total length of one unit cell is $W = W_L + W_M + W_R + W_S$.

We will define a slightly different formalism than in the previous section. Instead of a transfer matrix that connects wave function coefficients in the different regions, we will define a characteristic matrix that connects the wave functions on the two sides of a single region \cite{MasKro},
\begin{equation}
  \psi_i(x) = M_i \psi_i(x+W_i).
\end{equation}
The characteristic matrix for a single region is given by,
\begin{equation} \label{matrix_single}
  M_i = \frac{1}{\cos{\theta_i}}
  \begin{pmatrix}
    \cos{(k_{i} W_i + \theta_i)}   & -i f_i^{-1} \sin{k_{i} W_i} \\
    -i\ f_i \sin{k_{i} W_i}      & \cos{(k_{i} W_i - \theta_i)}
  \end{pmatrix},
\end{equation}
where $i=L,R,M,S$ and the potentials in the spacing regions are zero: $V_M=V_S=0$ and $A_M=A_S=0.$ The characteristic matrix for the whole structure is obtained by multiplying the individual matrices in the order as they appear in the structure $M = \prod_{i} M_i$.

For $N$ unit cells the total matrix becomes $M^N$ and we use a result from the theory of matrices, according to which the $N$th power of a unimodular matrix $M$ is ($u_N(\chi)\equiv u_N$),
\begin{equation} \label{matrix_total_N}
  M^N =
  \begin{pmatrix}
  M_{11}u_{N-1} - u_{N-2} & M_{12} u_{N-1}\\
  M_{21} u_{N-1} & M_{22}u_{N-1} - u_{N-2}
  \end{pmatrix},
\end{equation}
with $\chi=\frac{1}{2}TrM$ and $u_N$ is the Chebyshev polynomials of the second kind:
\begin{equation}
  u_N(\chi) = \frac{\sin[(N+1)\zeta]}{\sin\zeta},
\end{equation}
where $\zeta$ is the Bloch phase given by $\zeta = \arccos(\chi)$ \cite{MasKro}.

The total characteristic matrix gives us the relation between the wave functions just before and after the barrier structure,
\begin{equation} \label{matrix_relation0}
  \psi_{in}(0) = M^N \psi_{out}(NW).
\end{equation}
The transmission coefficient can be found by solving the system given by Eq. (\ref{matrix_relation0}),
\begin{eqnarray}
  1 + r &=& t e^{ik_0W} (M_{N11} + M_{N12} f_0 e^{i\theta_0}), \\
  f_0 (e^{i\theta_0} - r e^{-i\theta_0}) &=& t e^{ik_0W} (M_{N21} + M_{N22} f_0 e^{i\theta_0}),
\end{eqnarray}
which leads to,
\begin{equation}
  \label{transmission_general}
  t = \frac{2e^{-ik_0W} \cos{\theta_0}}{M_{N11}e^{-i\theta_0} + M_{N12}f_0 + M_{N21}f_0^{-1} + M_{N22}e^{i\theta_0}}.
\end{equation}
\begin{figure}[t]
  \centering
  \includegraphics[width=8.6cm]{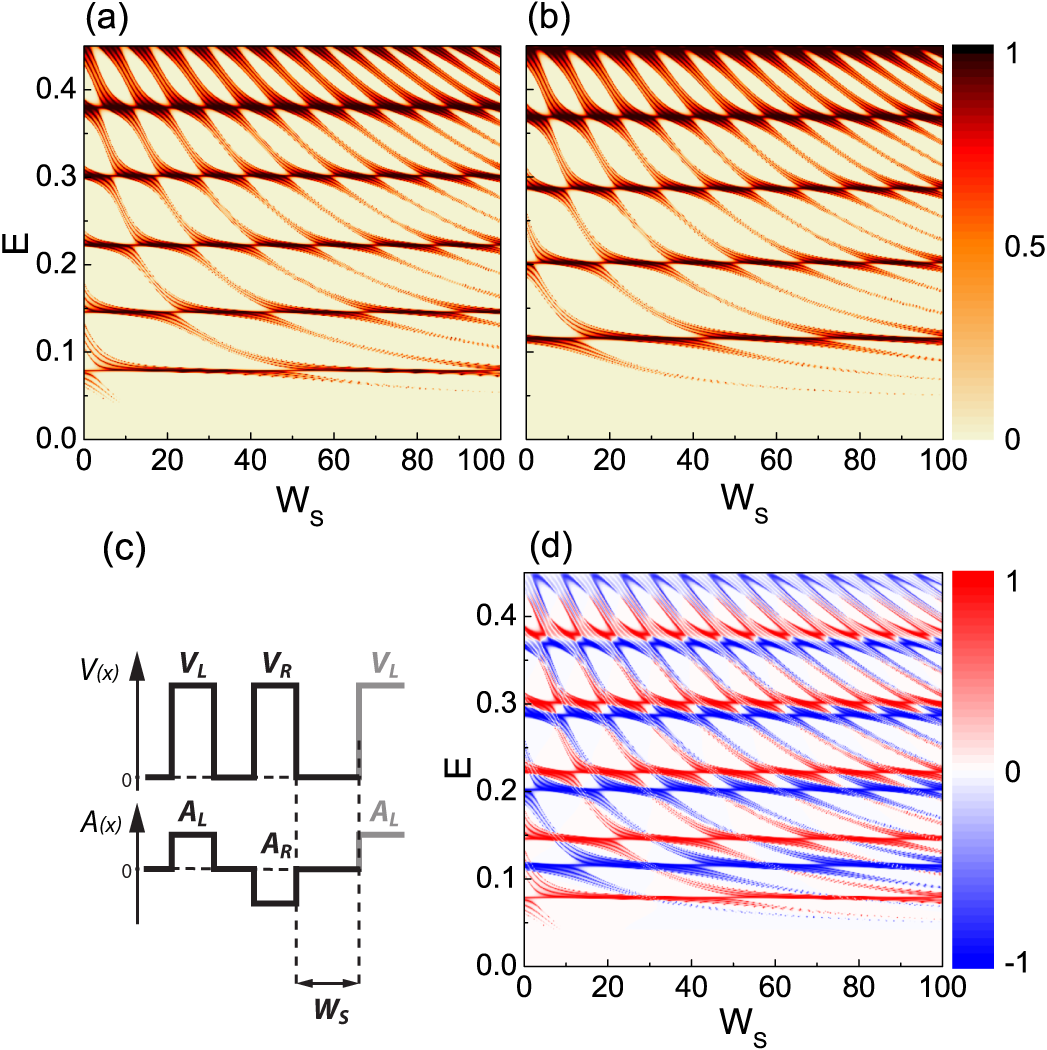}
  \caption{(Color online) Transmission probability at $k_y=0$ as a function of energy and the unit cell spacing $W_S$ for the (a) $K$ and (b) $K'$ valleys. (c) The barrier structure consists of a symmetric electric potential and an antisymmetric vector potential. (d) The difference $T_K - T_{K'}$. The parameters are: $V_L=V_R=0.2$, $A_L=-A_R=0.1$, $\mu=0.0425$, $W_L=W_M=W_R=10$.}
  \label{fig-L5-4}
\end{figure}
\emph{\textbf{Numerical results:}} As an example, we consider a system consisting of $N=5$ unit cells. The numerical results of transmission and energy states are presented in Figs. \ref{fig:B4_N5_transmission}(a) and (b) for the $K$ and $K'$ valleys respectively. The valley differences are not diminished with the addition of more barriers. Both the transmission and bound states are valley dependent which suggests that a superlattice of this kind of system will exhibit different bands for $K$ and $K'$. We also plot the transmission probability as a function of the energy $E$ and the unit cell spacing $W_{S}$.  As shown in Figs. \ref{fig-L5-4}(a), (b) and (d) with an increased number of barriers the difference between the transmission in $K$ and $K'$ is more pronounced as compared to the single cell case (Fig. \ref{fig:B4_transmission_Wvar}). The resonances are split into multiple peaks due to the presence of the additional unit cells. When adding one extra unit cell the new electron wave function can be written as a symmetric or antisymmetric combination and each transmission peak splits to two. As we add $N$ unit cells there are $N$ eigenstates that arise from linear combinations of the $N$ states of the independent unit cells and as a result the maximum of the transmission will split into $N$ peaks (e.g. in Fig. \ref{fig-L5-4} for $N=5$ we see that the resonances are split into $5$ peaks). As shown in Figs. \ref{fig-L5-4} (a) and (b) for some specific energies we have almost straight lines of full transmission. These energies correspond with the positions of maximum transmission shown in Figs. \ref{fig:B4_transmission_const_A} (a) and (b) for $k_{y}=0$ and they are given by full resonant transmission peaks described by Eq. (\ref{transmission_phase_max}).

\section{Superlattice}

Next, we take the same unit cell as described in the previous section, but consider the limit of $N \rightarrow \infty $ for a superlattice. The spectrum of the superlattice is obtained from the characteristic matrix of the unit cell as,
\begin{equation} \label{sl_spectrum_base}
  \cos(k_x W) = \frac{1}{2} Tr(M) = \frac{1}{2} (M_{11} + M_{22}).
\end{equation}
Now it is no longer meaningful to talk about transmission. Instead, we will study the energy band structure, the density of state and the conductivity.

If the spacing between unit cells $W_S$ is much smaller than the width of the other regions, i.e. $W_S \ll W_L,W_M,W_R$, the characteristic matrix for the cell spacing region can be replaced by an identity matrix. We will keep $W_M$ proportional to $W_L$ and $W_R$. The reason for this will be shown later. After making the substitution $M_S \rightarrow I$ and calculating the matrix for the whole structure, the matrix elements are used in Eq. (\ref{sl_spectrum_base}) to derive the implicit relation to obtain the energy spectrum of the superlattice,
\begin{eqnarray}
  \label{SL_spectrum}
  &\cos(k_x W) = \cos{\alpha_L}\cos{\alpha_M}\cos{\alpha_R} + \\*
  \nonumber
  &\gamma_\tau \sin{a_L}\sin{\alpha_M}\sin{\alpha_R} +
  \Omega_{L,M,R} + \Omega_{M,R,L} + \Omega_{R,L,M},
\end{eqnarray}
where,
\begin{equation} \label{gamma_SL}
  \gamma_\tau = \frac{\tau\mu}{k_{L} k_{M} k_{R}} \left[ A_L\mathcal{V}_{M,R} + A_M\mathcal{V}_{R,L} + A_R\mathcal{V}_{L,M} \right],
\end{equation}
\begin{equation}
  \Omega_{i,j,k} = \frac{-1}{k_{i} k_{j}} \left[ \varepsilon_i\varepsilon_j - \mu^2 - \kappa_i\kappa_j \right] \sin\alpha_i\sin\alpha_j\cos\alpha_k,
\end{equation}
and $\alpha_i = k_{i}W_i$, $\mathcal{V}_{i,j} = V_i - V_j$, $\varepsilon_i = E - V_i$, $\kappa_{i} = k_y + A_i$.

The spectrum of the superlattice depends on the valley index through $\gamma_\tau$. This valley dependent factor is only present if the superlattice has at least three potential regions. If we remove a region, e.g. $W_M = 0$, we have $\sin{\alpha_M} = 0$ and $\gamma_\tau$ is no longer a factor in the energy dispersion.

A mass term is required in order to have a non-zero $\gamma_\tau$. Valley dependent behavior of the superlattice also depends on the specific electric and vector potential values in the three regions. This potential relation is a bit more complicated, but Eq. (\ref{gamma_SL}) can also be represented in a determinant form,
\begin{equation} \label{gamma_SL_det}
  \gamma_\tau = \frac{\tau}{k_{xL} k_{xM} k_{xR}}
  \begin{vmatrix}
  A_L & -A_M & A_R \\
  V_L & V_M & V_R \\
  \mu & \mu & \mu
  \end{vmatrix}.
\end{equation}
If all the values in a row or column of the determinant are zero then the result is zero.

The previously derived relations are valid for any general case of three potentials, however we are interested in the specific case where the middle region corresponds to the space between two barriers. Therefore, in that region, the electric and vector potentials are zero: $V_M = 0$ and $A_M = 0$. This will not remove $\gamma_\tau$ completely because there is still a constant mass term in the second region. In this simpler case, the valley factor becomes,
\begin{equation} \label{gamma_SL_2}
  \gamma_\tau = \frac{\tau\mu}{k_{L} k_{M} k_{R}} \left( V_LA_R - V_RA_L \right),
\end{equation}
which is similar to the valley condition Eq. (\ref{gamma_barrier}) of the previously described barrier structure. We will consider a symmetric electric potential $V_L=V_R=V$ with an asymmetric vector potential $A_L=-A_R=A$. In that case the valley dependent factor becomes simply,
\begin{equation} \label{gamma_SL3_V_const}
  \gamma_\tau = \frac{-2\tau\mu}{k_{L} k_{M} k_{R}} VA.
\end{equation}
\begin{figure}[t]
  \centering
  \includegraphics[width=7.6cm]{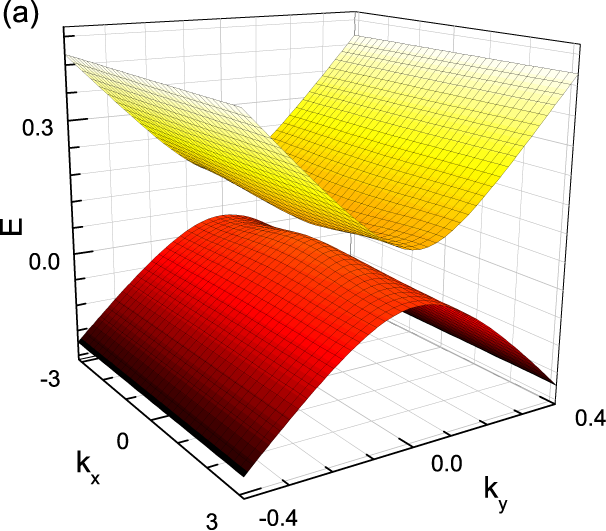}
  \includegraphics[width=7.6cm]{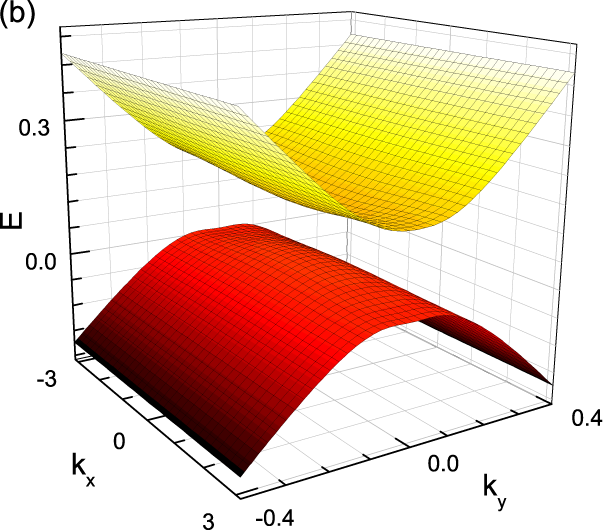}
  \caption{(Color online) Superlattice spectrum for the (a) $K$ and (b) $K'$ valleys. The parameters are: $V_L=V_R=0.16$, $A_L=-A_R=0.1$, $\mu=0.0425$, $W_L=W_M=W_R=10$, $W_S=2$.}
  \label{fig:SL4_bands_3D_V0.16}
\end{figure}
\begin{figure}[t]
  \centering
  \includegraphics[width=7.6cm]{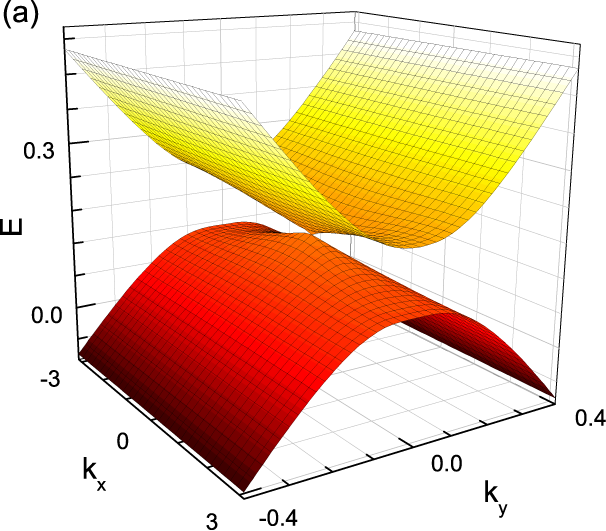}
  \includegraphics[width=7.6cm]{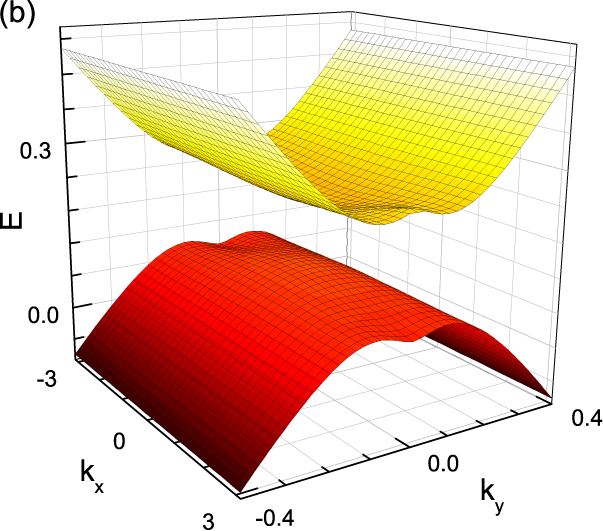}
  \caption{(Color online) Superlattice spectrum for the (a) $K$ and (b) $K'$ valleys. The parameters are: $V_L=V_R=0.29$, $A_L=-A_R=0.1$, $\mu=0.0425$, $W_L=W_M=W_R=10$, $W_S=2$.}
  \label{fig:SL4_bands_3D_V0.29}
\end{figure}
\begin{figure}[t]
  \centering
    \includegraphics[width=8.6cm]{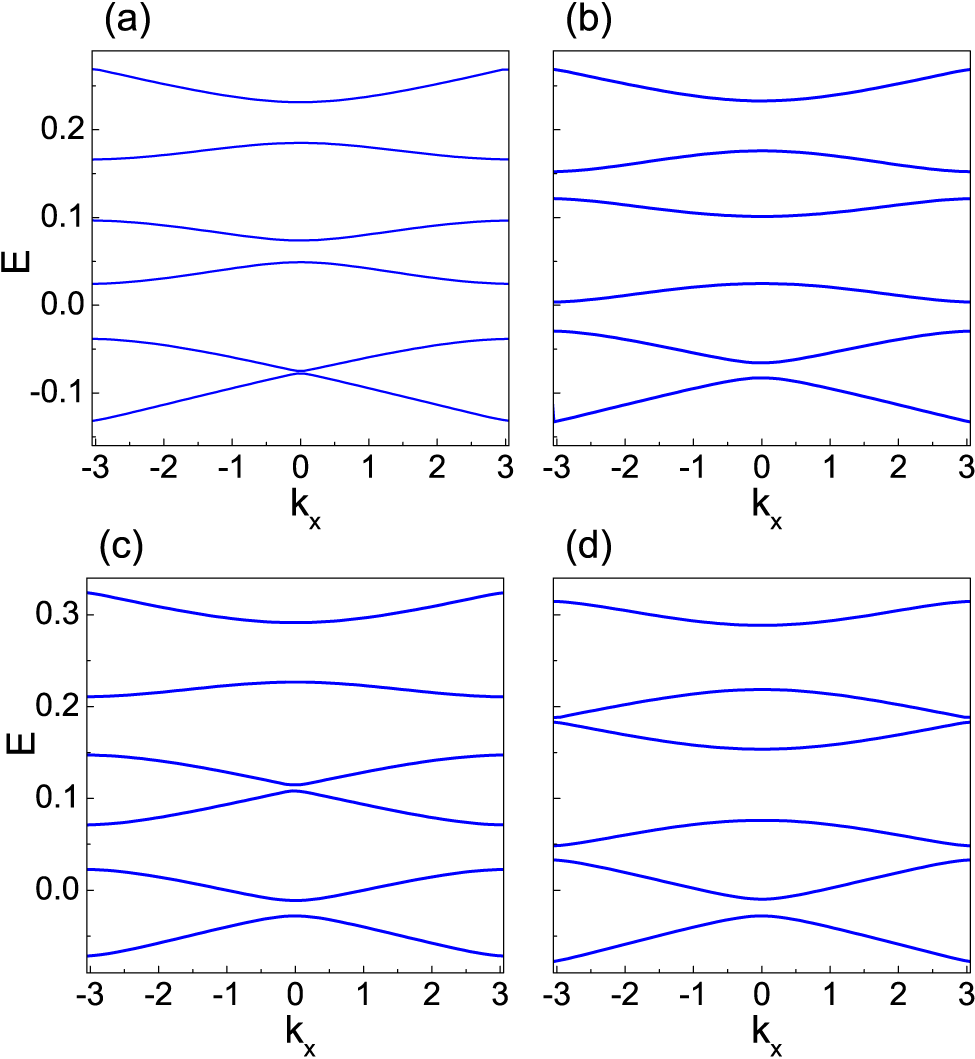}
    \caption{(Color online) Dispersion relation ($E$ vs $k_{x}$) for $k_y=0$ for the (a) $K$ and (b) $K'$ valleys with $V_L=V_R=0.16$ and for the (c) $K$ and (d) $K'$ valleys with $V_L=V_R=0.29$. The parameters are: $A_L=-A_R=0.1$, $\mu=0.0425$, $W_L=W_M=W_R=10$, $W_S=2$.}
    \label{figSLa}
\end{figure}

It is important to mention that in all cases the valley dependent behavior requires a real external magnetic field. The same could not be achieved with a strain generated field. The sign of the strain induced vector potential is different for each valley and it cancels out the previous valley differences. This is easy to show from the simplified valley factor given by Eq. (\ref{gamma_SL_2}). In the case of a strain induced vector potential the substitution $A_i \rightarrow \tau A_i$ would have to be introduced. That leads to,
\begin{eqnarray}
  \gamma_\tau = \frac{\tau^2\mu}{k_{xL} k_{xM} k_{xR}} \left( V_LA_R - V_RA_L \right),
\end{eqnarray}
and because $\tau^2=1$ the valley index vanishes from the energy dispersion relation. The same conclusion is true for the general three region valley factor from Eq. (\ref{gamma_SL}).

\emph{\textbf{Density of states (DOS):}} The number of k-states per unit energy is given by
\begin{equation}\label{8-D1}
    \displaystyle{D(E) = \frac{1}{(2 \pi)^{2}}\sum_{n}\int{dk_xdk_{y}\,\delta(E - E_{n}(k_x,k_{y}))}}.
\end{equation}
To calculate the DOS numerically we introduce a Gaussian broadening,
\begin{equation}\label{8-D5}
    \displaystyle{\delta(E-E_{n}(k_x,k_{y})) \rightarrow \frac{1}{\Gamma \sqrt{\pi}}\exp{\left[-\frac{(E -
    E_{n}(k_x,k_{y}))^2}{\Gamma^{2}}\right]}}.
\end{equation}

\emph{\textbf{Conductivity:}} For elastic scattering the diffusive conductivity $\sigma_{ij}$ is given by
\begin{equation}\label{8-C1}
    \sigma_{ij} = \frac{e^{2}}{4\pi^{2}k_BT}\sum_{n}\int dk_xdk_{y}\tau_m\,v_{ni}v_{nj} f_{n{\bf k}} (1-f_{n{\bf k}}).
\end{equation}
Here $T$ is the temperature, $v_{ni}=\p E_{n}/\p k_{i}$  the electron velocity, $f_{n{\bf k}}$ the Fermi-Dirac function, and $\tau_m$ the momentum relaxation time. For low temperatures we assume that $\tau_m$ is approximately constant, evaluated at the Fermi level ($\tau_m \approx \tau_F$), and replace the product $ f_{n{\bf k}} (1-f_{n{\bf k}})/k_BT$ by the delta function $\delta(E -E_{n}(k_x,k_{y}))$.

\emph{\textbf{Numerical results:}} The energy bands of the superlattice are calculated numerically based on the total transfer matrix and the energy dispersion relation Eq. (\ref{sl_spectrum_base}).

As mentioned previously, we will only consider a symmetric electric potential with an antisymmetric vector potential. The different band structures for $K$ and $K'$ valleys with this kind of potential function are shown in Figs. \ref{fig:SL4_bands_3D_V0.16} and \ref{fig:SL4_bands_3D_V0.29} for different values of the electric potential. In Fig. \ref{figSLa} we plotted the energy dispersion versus $k_{x}$ with a fixed $k_{y} = 0$. In both cases the dispersion relation is plotted for two values of the electric potential in order to show the large influence of the potential on the superlattice band structure.

\begin{figure}[t]
\begin{center}
\includegraphics[width=8.6cm]{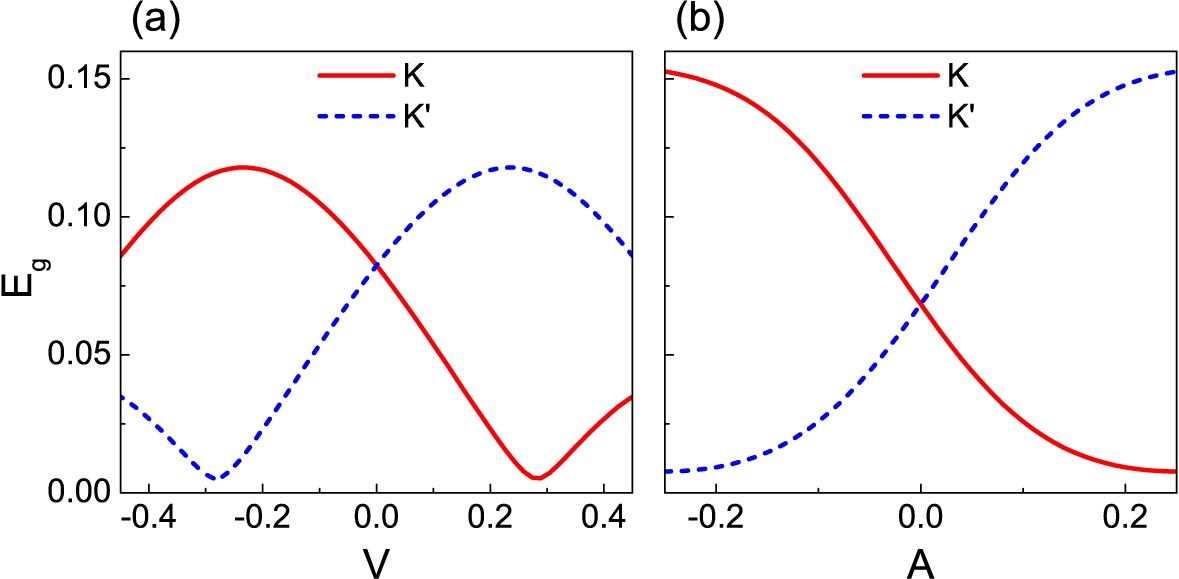}
\caption{(Color online) Band gap versus the electric and vector potentials. (a) The electric potential is tuned so that $V_L=V_R=V$, while the vector potential is taken constant with $A_L=-A_R=0.1$. (b) The vector potential is tuned so that $A_L=-A_R=A$, while the vector potential is taken constant with $V_L=V_R=0.2$. The rest of the SL parameters are the same for (a) and (b): $\mu=0.0425$, $W_L=W_M=W_R=10$, $W_S=2$.} \label{fig:SL4_band_gap}
\end{center}
\end{figure}

Along with the general band structure, the band gap depends on the valley index. Its value can be tuned by adjusting the electric and vector potentials, as shown in Fig. \ref{fig:SL4_band_gap}. At the point where either the electric or vector potential is zero, the band gap is identical for both valleys which is expected given the valley factor $\gamma_\tau$(\ref{gamma_SL3_V_const}). The energy gap functions for the valleys are reversed with regards to the zero potential point:
\begin{eqnarray}
  E_g(\tau, V, A) = E_g(-\tau, -V, A), \\
  E_g(\tau, V, A) = E_g(-\tau, V, -A),
\end{eqnarray}
which is also obvious from the valley factor expression.

It is possible to adjust the values so that one valley has a narrow band gap, while the other has a wide one, as is shown in Fig. \ref{fig:SL4_bands_3D_V0.29}. This case depicts the band structure at the point $V=0.29$ from the gap function in Fig. \ref{fig:SL4_band_gap}. It is particularly convenient that the valley behavior of the superlattice can easily be reversed by flipping the electric potential from $V$ to $-V$ or $A$ to $-A$.
\begin{figure}[b]
\begin{center}
\includegraphics[width=5.4cm]{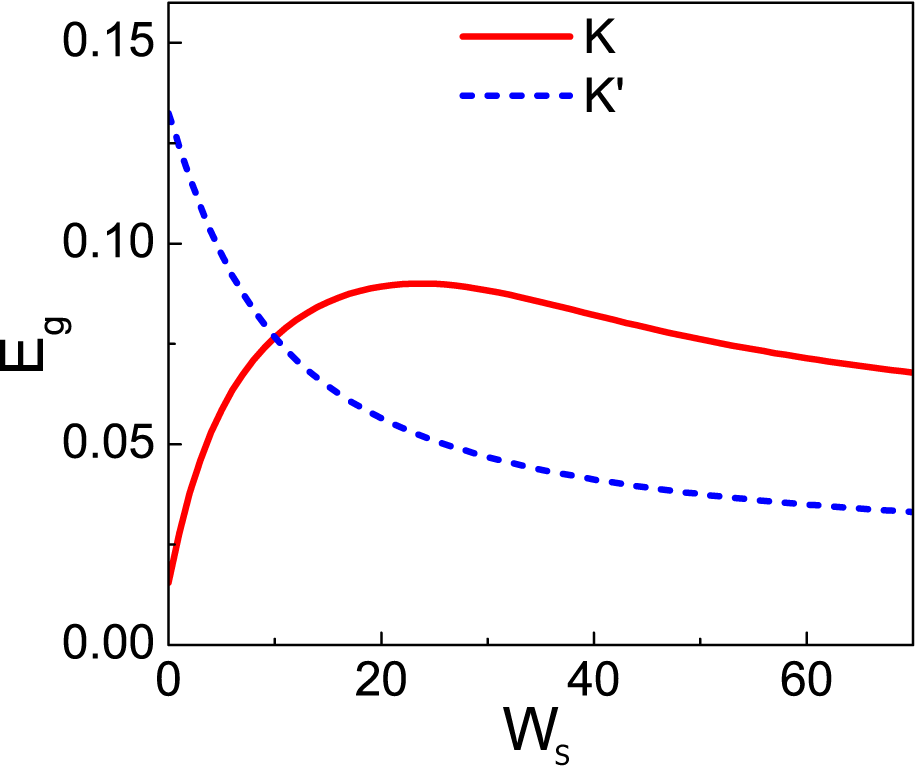}
\caption{(Color online) Band gap versus the length of the unit cell spacing $W_S$ for the energy spectrum of the $K$ and $K'$ valleys. The SL parameters are: $V_L=V_R=0.16$, $A_L=-A_R=0.1$, $\mu=0.0425$, $W_L=W_M=W_R=10$.}
\label{fig:SL4_band_gap_W}
\end{center}
\end{figure}
\begin{figure}[t]
\begin{center}
\includegraphics[width=6.0cm]{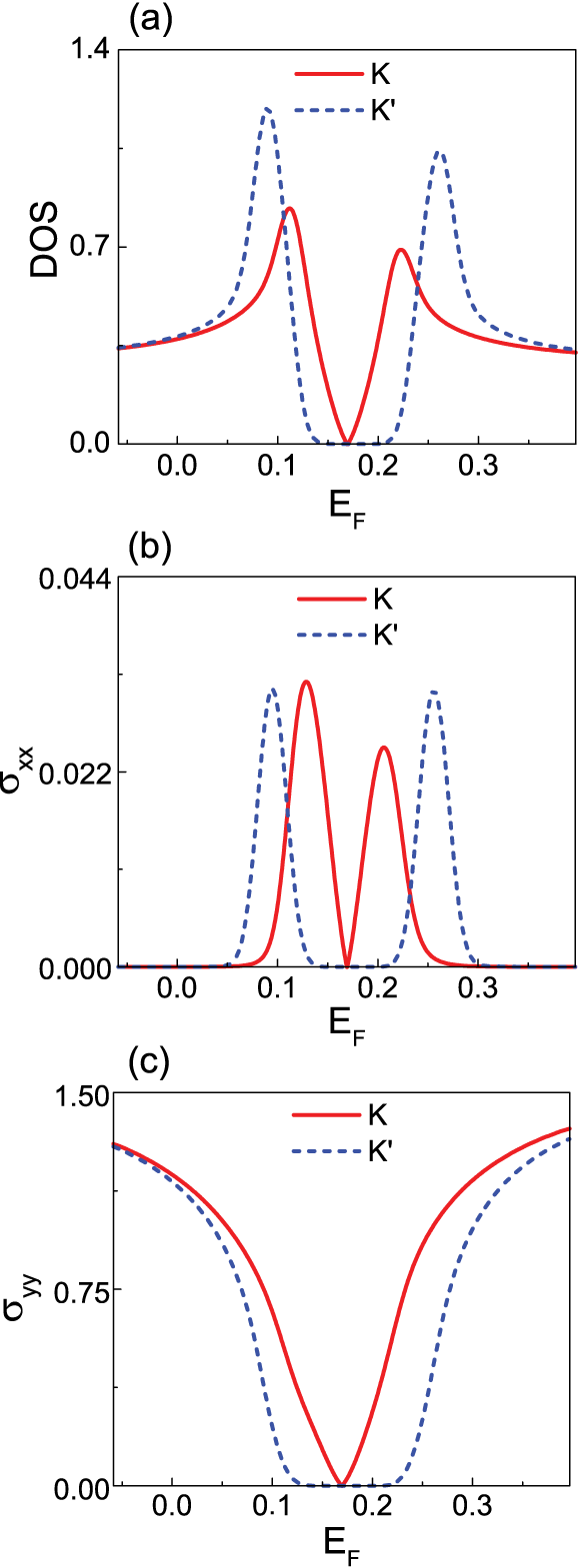}
\caption{(Color online) (a) Density of states (b) and (c) conductivities vs Fermi energy for $K$ and $K'$ valleys. The parameters are: $V_L=V_R=0.29$, $A_L=-A_R=0.1$, $\mu=0.0425$, $W_L=W_M=W_R=10$, $W_S=2$.} \label{fig:SL4_DOS}
\end{center}
\end{figure}

The energy spectrum and band gap of the superlattice also depends on the spacing between the barriers. This dependence is plotted in Fig. \ref{fig:SL4_band_gap_W}. For $W_S = 10$ there is no difference between $K$ and $K'$. At that point the widths of all four regions are identical, which makes both the electric and vector potential functions symmetric from the point of view of the entire superlattice. Note, the potential functions are not symmetric locally in a single unit cell, but only globally. On the other hand, the valley difference is most pronounced at $W_S=0$, because we have a globally symmetric electric potential with a globally antisymmetric vector potential.

The numerical results for the DOS and the conductivities $\sigma_{xx}$ and $\sigma_{yy}$ for $V=0.29$ are plotted in Fig. \ref{fig:SL4_DOS}. These results show that the valley differences are localized to energy values around the band gap. The largest valley differences are present at the center of the band structure, while at energy values further away from the band gap $K$ and $K'$ converge to the same values, both in the case of the DOS and the conductivities. It should be noted that $\sigma_{yy} \gg \sigma_{xx}$ and that $\sigma_{xx}$ goes to zero at energy values further away from the band gap, which means that the electrons have zero group velocity in the $x$ direction.

\section{Conclusions}
We proposed a model consisting of electric and vector potentials for a massive Dirac electron in graphene. The model is essentially a series of very high and very narrow magnetic $\delta$-function barriers alternating in signs, plus electric potential barriers. The transmission through such a series of barriers was obtained using transfer matrix methods.

First, we showed that although a single barrier presents the same transmission and reflection probabilities regardless of the valley index, electrons from different valleys are in fact reflected with a different phase. When a second barrier is added, this valley phase difference will affect the resonant peaks that occur between the two barriers. We show that the transmission probability for this double barrier structure is very different for electrons in different Dirac points. This kind of structure can be formed by using two ferromagnetic stripes (with in-plane magnetization), which can also be used as electric gates. A mass term must also be present in order that this structure acts as a valley filter. A gap in the electronic spectrum can be induced either by a substrate or by electron electron interactions.

We considered a few unit cells (five) and showed that the valley filtering behavior of the structure is robust. This enabled us to extend the calculation to a superlattice for which the unit cell corresponds to two stripe ferromagnetic gates. We show that depending on the configuration of the magnetization and electric potential of the gates, different band structures appear for electrons in the two Dirac points. The generated band gap is also valley dependent and can be easily tuned by changing the electric potential or the magnetization. Even more, the behavior of electrons in the two valleys can be switched by flipping the sign of either potential ($V \rightarrow -V$ or $A \rightarrow -A$).

From an experimental point of view the valley filters and polarizers could be easily realized as superlattices (or even finite number of unit cells) made of ferromagnetic stripes all magnetized in-plane and in the same direction but with alternating electric potentials. Alternatively, although less feasible experimentally, one can also obtain valley filtering behavior for alternating magnetization but with the same electric potentials. An essential ingredient is the presence of a finite gap in the electronic spectrum, which is nowadays routinely achieved in graphene deposited on BN substrates.

\begin{acknowledgments}
This work was supported by the European Science Foundation (ESF) under the EUROCORES Program EuroGRAPHENE within the project CONGRAN and the Flemish Science Foundation (FWO-Vl).
\end{acknowledgments}

\end{document}